\documentclass[a4paper, a4paperamsfonts, amssymb, amsmath,showkeys, nofootinbib, twoside]{revtex4-2}
\usepackage[english]{babel}
\usepackage[utf8]{inputenc}
\usepackage[colorinlistoftodos, color=green!40, prependcaption]{todonotes}

\usepackage[pdftex, pdftitle={Article}, pdfauthor={Author}]{hyperref} 
\usepackage[T1]{fontenc}
\usepackage{lmodern}
\usepackage{amsmath}
\usepackage{amssymb}
\usepackage{xcolor}
\usepackage{graphicx}
\usepackage{caption}
\usepackage{subcaption}
\usepackage{float}
\usepackage{esint}
\usepackage{dcolumn}
\usepackage{babel}
\usepackage{csquotes}
\usepackage{color}
\usepackage{slashed}
\usepackage{simplewick}
\usepackage{hyperref}
\hypersetup{
    colorlinks=true,
    citecolor=blue,
    filecolor=green,
    linkcolor=purple,
    urlcolor=red,
}

\usepackage{slashed}

\usepackage{hyperref}
\hypersetup{colorlinks,breaklinks,
			citecolor=[rgb]{0,0.0,1.0},
            urlcolor=[rgb]{0.0,0.0,1.0},
            linkcolor=[rgb]{0,0.5,0.9}}

\begin{document}

\title{Einstein-Yang-Mills Regular Black Holes in Rainbow Gravity}


\author{Celio R. Muniz}
\email{celio.muniz@uece.br}
\affiliation{Universidade Estadual do Cear\'a (UECE), Faculdade de Educa\c{c}\~ao, Ci\^encias e Letras de Iguatu, Av. D\'ario Rabelo s/n, Iguatu - CE, 63.500-00 - Brasil.}
\author{Francisco Bento Lustosa}
\email{chico.lustosa@uece.br}
\affiliation{Universidade Estadual do Cear\'a (UECE), Faculdade de Educa\c{c}\~ao, Ci\^encias e Letras de Iguatu, Av. D\'ario Rabelo s/n, Iguatu - CE, 63.500-00 - Brasil.}
\author{Takol Tangphati}
\email{takoltang@gmail.com}
\affiliation{School of Science, Walailak University,  \\
Thasala, Nakhon Si Thammarat, 80160, Thailand}
\affiliation{Research Center for Theoretical Simulation and Applied Research in Bioscience and Sensing, Walailak University, Thasala, Nakhon Si Thammarat 80160, Thailand}

\date{\today}
\begin{abstract}
In this work, we investigate regular black hole solutions in nonminimal Einstein-Yang-Mills theory modified by Rainbow Gravity, focusing on the impact of quantum gravity effects on their thermodynamics, particle emission, energy conditions, curvature, and shadow formation. We find that the rainbow parameter $\lambda$ alters Hawking's temperature, entropy, and specific heat, leading to modified phase transitions and the possible formation of remnants. We calculate the graybody factor demonstrating that particle emission is enhanced with increasing $\lambda$, reflecting the behavior of the temperature and confirming the impact of the rainbow parameter on the evaporation process. Energy conditions are violated inside the black hole, with violations intensifying for larger $\lambda$. We also show that Rainbow Gravity mitigates singularity formation by softening the curvature near the origin, contributing to the regularity of the solution. Finally, we study the black hole shadow and demonstrate that its radius decreases as quantum gravity effects strengthen, suggesting potential observational tests for Rainbow Gravity. These results highlight the role of Rainbow Gravity in modifying black hole physics and provides a framework for exploring quantum gravitational corrections in astrophysical scenarios.
\end{abstract}

\keywords{Regular black holes, Einstein-Yang-Mills, Rainbow Gravity, Remnants}

\maketitle
\vspace{-0.5cm}
\section{Introduction}

Black holes (BHs) continue to be the prime laboratories to test our understanding of the intersection of gravitational and quantum phenomena \cite{Addazi2021quantumgravphenom, AlvesBatista_2025}. Since the 1970's, the study of BHs has been an incredible theoretical tool to test the consistency of our theories and drive the advancement of new ideas, most notably in the pursuit of a quantum theory of gravity \cite{Hawking1983,Maldacena2023}. More recently, this theoretical effort has been supported by an increasing amount of observational data \cite{Amelino-Camelia2013}. From the analysis of X and gamma ray bursts \cite{Kaaret:2017tcn}, continuing through the detection of gravitational waves \cite{LIGOScientific:2020ibl} and to the detailed observation of black hole shadows \cite{EventHorizonTelescope:2022wkp} we have nowadays several ways of using BH physics to test the limits of our theories \cite{Vagnozzi:2022moj}. 

In this context, it has been of great interest in the gravitation community to develop different BH models with possible novel features that could produce interesting phenomenological signals while solving some of our fundamental theoretical issues. One such issue is the singularity problem of General Relativity (GR) (for a recent review of the issue and possible solutions see \cite{Ashtekar2022}). However, there are several known ways to go around this problem with minor modifications to standard GR, such as considering quantum effects near the singularity \cite{Ashtekar:2018cay}, non-linear theories of gravity and electromagnetism \cite{Ayon-Beato:1998hmi}, exotic matter sources \cite{Dymnikova1992, Alencar_2023} and many others \cite{Simpson_2019, Maluf2019, Maluf2022, Estrada_2025}. Another approach consists in modifying gravity through nonminimal couplings, assuming that some quantum field corrections are expected at sufficiently high energies \cite{Muller-Hoissen_1988,Faraoni:1998qx}. This can be motivated not only by the singularity problem but also by the quest for a better understanding of the dark sector of our universe \cite{ji2021wavedarkmatternonminimally,sankharva2022nonminimalcoupledDM,Lebedev_2023}. Some nonminimally coupled models lead to exact solutions of regular black holes that have been studied in the literature \cite{BalakinZayats2007, BalakinLemos2008}. Considering the current challenges that our cosmological and particle standard models face, a better understanding of possible extensions of the interaction between matter fields and gravity in different contexts is extremely relevant. In this work we will explore a model based on a nonminimal Einstein–Yang-Mills theory with $SU(2)$ symmetry and a Wu-Yang-type ansatz that was studied in \cite{BalakinZayats2007, BalakinLemos2016}.  

Beyond considering extensions to known theories, the pursuit of a fundamental quantum theory of gravity continues in many fronts \cite{Bambi2024}. In the context of Loop Quantum Gravity, for example, there is a clear solution for the singularity problem and several cosmological models have been developed \cite{Agullo:2016tjh}, but comparing them with actual observables and making them competitive against the current standard paradigm of cosmology is still a challenge \cite{Delgado_2024}. A common thread among different Quantum Gravity (QG) approaches is the breaking of the Lorentz symmetry at a fundamental level that can lead to possible observational constraints \cite{Alfaro2005, Magueijo2005stringtheory,Ellis2002}. This possibility has lead to a wide variety of both phenomenological models and observational bounds \cite{Addazi2021quantumgravphenom, AlvesBatista_2025}. Exploring this common feature of QG theories, an approach that has been gaining attention in the recent literature is based on a possible deformation or extension of special relativity, Doubly Special Relativity (DSR) \cite{Magueijo2003, amelino-camelia2002}. The extension of this framework to curved space-times and gravitational fields is now called the Rainbow Gravity theory \cite{Magueijo_2004}. 

In the framework of DSR modified dispersion relations have to be introduced to account for the energy dependence of the transformation of inertial frames with the quanta of energy when they are near the Planck Energy ($E_{Pl}$). This leads to a set of nonlinear transformations in momentum space that can be translated into a modified space-time metric that will also depend on the energy of the quanta and the Planck scale, so a given metric will now be described by a spectrum, hence, gravity's rainbow. This modified space-time picture will necessarily give rise to new phenomenological features that have been explored in a wide range different scenarios including black hole physics \cite{Gallan2006}, neutron stars \cite{Hendi_2016}, cosmological perturbations \cite{Wang2015, Leyva_2023} and other interesting effects involving quantum fields \cite{Bezerra2017, Bezerra_2017, BEZERRA2019, Alencar2020, Bakke2022, Furtado_2022}. An active line of research is the study of BH thermodynamics and the evaporation of BHs in the context of Rainbow Gravity \cite{Li2007,Ahmed2014,Feng2017-thermodynamicphasetransitioninraibow,Hendi2016,Ali2015,JUNIOR2020,Morais2022}. This can have various important consequences to the understanding of the formation of black hole remnants and information loss in the context of Rainbow Gravity \cite{Chen2015}. Independent of the singularity problem or the fundamental theory of QG we are dealing with, the study of BH thermodynamics can serve as theoretical consistency check on different types of models that can lead to useful information about fundamental physics. 

Our analysis contributes to the ongoing debate regarding the existence and stability of remnants in Rainbow Gravity \cite{Ahmed2014, JUNIOR2020,Morais2022}. The possible existence of BH remnants has a wide range of consequences, from solutions to the information loss paradox \cite{Chen2015} to possible Dark Matter candidates being generated by the evaporation of primordial BHs into stable remnants \cite{Yang2020pbhasDM}. However, in the context of the model we will be examining a few points have to be highlighted. Magnetically charged BHs have an unusual evaporation behavior even in standard Einstein-Maxwell theory and its final fate continuous to be a subject of investigation \cite{Kim2004,Maldacena2021}. In the context of 4D Einstein-Yang-Mills theory the stability of BH solutions has been questioned \cite{VOLKOV1995438} and discussed in connection with violations of the no hair theorem \cite{Ashtekar_2001}. The nonminimally coupled solution first found in \cite{BalakinZayats2007} and further discussed \cite{BalakinLemos2016} is regular in the gravitational field but the gauge field remains singular at the origin, indicating that evaporation will necessarily go through a phase where the magnetic field dominates over gravity and the final fate of the sytstem will depend on the fundamental properties of the theory regarding magnetic monopoles \cite{Kim2004,Maldacena2021}. With all that taken into account, our work consists of studying the effects of modifications of the dispersion relations in the structure of space-time itself through the calculation of both thermodynamical quantities and geometrical properties of the BH system in the nonmiminmally coupled Einstein-Yang-Mills theory with the Wu-Yang type magnetic monopole ansatz \cite{BalakinZayats2007}.

The aforementioned debate on black hole remnants in Rainbow Gravity has been mainly focused on possible connections with the Generalized Uncertainty Principle (GUP) and its correct use depending on the different choices for the rainbow functions that determine the modification of the dispersion relations \cite{Gallan2006,Morais2022}. A less discussed topic is the running of the gravitational constant $G$ and how it should be modeled \cite{Donoghue2012}, which can be argued to be a necessary feature of any Rainbow Gravity proposal in order to guarantee that different observers agree with the scale of energy in which Lorentz breaking should occur \cite{Magueijo_2004,NILSSON2017115}. Although some works have incorporated an energy dependent gravitational coupling explicitly \cite{JUNIOR2020,GIM2019122}, we are not aware of a detailed study considering the impact of an energy dependent $G(E)$ on the formation of remnants in the context of Rainbow Gravity. In this work we will considered a specific combination of rainbow functions and gravitational coupling running that was first suggested in \cite{Magueijo2002,Magueijo_2004}.

In this paper we will provide the thermodynamical analysis of the Black Hole solution with the nonminimal coupling with a Yang-Mills gauge field in the context of Rainbow Gravity in order to stablish the stability conditions and analyze its possible evolution towards the formation of remnants. Due to the nonminimal coupling, the model is already regular before the introduction of the rainbow functions.  We provide a brief discussion on what can and cannot be said about the formation of remnants from our models and then make use of the graybody factor calculations to corroborate our findings from the semiclassical thermodynamical quantities \cite{Al-Badawi:2023emj, Calza:2025whq}. We will then study what is the effect of the rainbow parameters in the violation of the energy conditions near the singularity. We will also show how the Ricci and Kretschmann scalars are affected by the introduction of the rainbow functions. Finally, pointing towards future phenomenological studies, we provide the effect of the quantum gravity parameter on the shadow size of our BH solution \cite{PERLICK20221,Liu2019pov, Kimet2021,Jiang_2024,LIU2024139052}. We observe a considerable variation in shadow size due to the introduction of the rainbow gravity modification, indicating the necessity for a more detailed and complete study to compare our results with observations. Introduction of rotation through Newman-Janis algorithm and a calculation on other effects of the shadow will be subject of future investigations that might lead to concrete tests of our models and put further constraints on the form of rainbow gravity's functions \cite{Jiang_2024, LIU2024139052}. 

The work is organized as follows. In the next section we will present our model, introducing the modifications due to Rainbow Gravity into the Einstein-Yang-Mills Black Hole solution of \cite{BalakinLemos2016}. In Section \ref{thermodynamics-section} we will calculate the related thermodynamic quantities, presenting our analysis of the temperature, entropy and specific heat. In Section \ref{gray}, we investigate the emission of massless particles by calculating the graybody factors. We proceed to the study of the energy and curvature behaviors of our solution in Section \ref{energyandcurvatures}, showing how the singularity avoidance is affected by the modifications introduced by Rainbow Gravity. Finally, in \ref{shadows} we calculate the radius of the photon sphere as a function of the rainbow metric function parameter indicating the path for a phenomenological study of our model. In the Conclusions we discuss our results and its possible implications and extensions. 
\section{Einstein-Yang-Mills regular black holes in Rainbow Gravity} \label{section1}

Nonminimally coupled Einstein-Yang-Mills theories can be motivated by the quest for a quantum theory of gravity through the addition of higher order terms in the Einstein-Hilbert action, as it was shown that this type of theory can arise from a dimensional reduction from an Gauss-Bonet type of action in 5 dimensions \cite{Muller-Hoissen_1988}. It was first shown in \cite{BalakinZayats2007} and further studied in \cite{BalakinLemos2016} that in this context and with a certain combination of parameters one can obtain regular black hole solutions. Since this seminal works, this solution has been studied from different perspectives highlighting its potential for testing modifications of standard GR \cite{Jawad2018cdh,Liu2019pov,Kimet2021,Al-Badawi:2023emj,Pu_2023}. 

The added assumption that fundamental Lorentz covariance emerges for low energy quanta within the context of Rainbow Gravity provides further modifications of the space-time structure that could potentially lead to deviations from previously calculated phenomenological properties of this black hole solution. We will briefly describe how the original Einstein-Yang-Mills nonminimally coupled solution was obtained, introduce the modified rainbow metric and define the specific phenomenological functions we will study, thus arriving at the necessary metric function that will be used in the following sections.
\vspace{-0.5cm}
\subsection{The Non-minimal Einstein-Yang-Mills Black Hole}

The nonminimal Einstein-Yang Mills theory action is given by \cite{BalakinZayats2007,BalakinLemos2016}
\begin{equation}
    S_{NMEYM} = \int d^4x \sqrt{-g}\biggl\{ \frac{R}{8\pi G c^{-4}}+ F_{ik}^{(a)}F^{ik(a)}  + \frac{1}{2}\mathcal{R}^{ikmn}F_{ik}^{(a)}F_{mn}^{(a)}  \biggr\},
\end{equation}
where $g$ is the determinant of the metric, $R$ is the usual Ricci scalar and  $F_{ik}^{(a)}$ are components of the Yang-Mills field originally described by a triplet of vector potentials $A_m^{(a)}$. The nonminimal susceptibility tensor $\mathcal{R}^{ikmn}$ is defined as
\begin{eqnarray}
        \mathcal{R}^{ikmn} \equiv \frac{q_1}{2}R(g^{im}g^{kn} - g^{in}g^{km}) \\
    + \frac{q_2}{2}(R^{im}g^{kn}- R^{in}g^{km}+R^{kn}g^{im}- R^{km}g^{in}) + q_3R^{ikmn},
\end{eqnarray}
where the phenomenological parameters $q_1,q_2$ and $q_3$ characterize the nonminmal coupling and $R^{ik}$ and $R^{ikmn}$ are the usual Ricci and Riemann tensors. Assuming an spherically symmetric metric in the general form of
\begin{equation}
    ds^2 = F(r) dt^2 - dr^2F(r)^{-1} - r^2d\Omega^2
\end{equation}
and the gauge-field with a Wu-Yang-type ansatz for the potential $A_m^{(a)}$ \cite{BalakinZayats2007}, we can obtain the nonminimal gravitational field equations for the metric function $F(r)$ of the form
    \begin{align}
\begin{split}\label{eq:1}
 \frac{1-F(r)}{r^2} - \frac{F'(r)}{r} ={}& \frac{8\pi G c^{-4}\nu^2 }{r^4} \times \\ 
 &\times\left[ \frac{1}{2}-q_1 \frac{F'(r)}{r} + (13 q_1 +4q_2 + q_3)\frac{F(r)}{r^2}-\frac{q_1 + q_2 + q_3}{r^2}\right],
\end{split}\\
\begin{split}\label{eq:2}
    \frac{1-F(r)}{r^2} - \frac{F'(r)}{r} ={}& \frac{8\pi G c^{-4}\nu^2 }{r^4}\times\\
         &\times\left[ \frac{1}{2}-q_1 \frac{F'(r)}{r} - (7 q_1 +4q_2 + q_3)\frac{F(r)}{r^2}-\frac{q_1 + q_2 + q_3}{r^2}  \right],
\end{split}\\
\begin{split} \label{eq:3}
    \frac{F'(r)}{r} + \frac{F''(r)}{2} ={}& \frac{8\pi G c^{-4}\nu^2 }{r^4} \times\\
        &\times \left[  \frac{1}{2}-q_1 \frac{F''(r)}{2} - (7 q_1 +4q_2 + q_3)\left( \frac{F'(r)}{r} - \frac{2F(r)}{r^2}\right) + 2\frac{(q_1 + q_2 + q_3)}{r^2} \right],
\end{split}     
\end{align}
where $\nu$ is a magnetic positive parameter. Equations \ref{eq:1} and \ref{eq:2} coincide when $10q_1 + 4q_2 +q_3 = 0$. Searching for regular solutions where $F(0) = 1$ and $F'(0)=0$ one also finds that \ref{eq:1} is compatible with these conditions when $4q_1 + q2 = 0$. 
This leads to the relations
\begin{eqnarray}
    q_1 \equiv -q, \ q_2 = 4q, \ q_3 = -6q,
\end{eqnarray}
and one assumes $q>0$. Using $GQ^2\equiv4\pi c^{-4}\nu^2$, we can rewrite the master equation for the gravitational field (3) as
\begin{equation}
    \frac{d}{dr}\left[ r(F(r) - 1)\left( 1 + \frac{2GQ^2q}{r^4}\right) \right] = - \frac{GQ^2}{r^2}.
\end{equation}
The solution for this equation can than be written as
\begin{equation}
    F(r)=1+\frac{r^4}{r^4+2 q G Q^2}\left(-\frac{2GM}{r}+\frac{GQ^2}{r^2}\right),
    \label{F(r)}
\end{equation}
where we have imposed the requirement of returning to the magnetically charged black hole solution in the limit $q \rightarrow 0$. We kept the gravitational constant $G$ explicit for reasons that will become clear in the next section. 

\subsection{Modified metric with Rainbow Gravity}

In the context of Rainbow Gravity, one assumes that Lorentz symmetry will be violated for fundamental particles that reach a high enough energy $E$ modifying its dispersion relations in momentum space. In \cite{Magueijo_2004} proposed a simple way of translating this modification from momentum space into metric space through the use of two phenomenological functions $f$ and $g$. Considering the general modified dispersion relation in the form
\begin{equation}
    E^2 f(\epsilon)^2 - p^2 g^2(\epsilon) = m^2,
\end{equation}
where $\epsilon=E/E_P$, we can obtain the modified \textit{ansatz} for the spherically symmetric metric as
\begin{equation}
ds^2=-\frac{F(r)}{f^2(\epsilon)}dt^2+\frac{dr^2}{F(r)g^2(\epsilon)}+\frac{r^2}{g^2(\epsilon)}d\Omega^2.
\end{equation}
We have now an energy dependent metric that will carry effects of the modified dispersion relations into the dynamics of space-time.

We can replace this ansatz on the equations of motion from the last subsection and obtain modified solutions by making the following changes in variables \cite{Magueijo_2004}
\begin{equation}
   \tilde{t} = \frac{t}{f(\epsilon)}, \ \tilde{r} = \frac{r}{g(\epsilon)}, \ d\tilde{t}= \frac{dt}{f(\epsilon)}, \ d\tilde{r} = \frac{dr}{g(\epsilon)}, \tilde{G}= \frac{G}{g(\epsilon)},
\end{equation}
where we have to introduce a modified (running) gravitational constant that reflects the expectation that the gravitational coupling itself will depend on the energy scale. It is important to note that some recent works have not considered a running gravitational constant in their analysis which might have consequences to their results regarding the existence of remnants \cite{Morais2022}. However, it must also be observed that there is no universal agreed upon definition for the running of the gravitational constant from effective field theory methods \cite{Donoghue2012}. Considering that, we follow the reasoning presented in \cite{Magueijo_2004} where an energy dependence on the constant of integration for the Schwarzschild metric is required for consistency of the solutions which translates into an energy dependence in the Newton constant of the form $ \tilde{G}= \frac{G}{g(\epsilon)}$.

We are now able to determine that making a change of variables is sufficient to obtain the solution to the modified field equations. We will consider the specific phenomenological rainbow functions \cite{Magueijo2002}
\begin{equation}\label{g-rainbow}
    f(\epsilon)=g(\epsilon)=\frac{1}{1+\lambda\epsilon},
    \end{equation}
where $\lambda$ is the rainbow parameter. We can write our new metric function in terms of the modified variables as
\begin{equation}
    F(\tilde{r})=1+\frac{\tilde{r}^4}{\tilde{r}^4+2 q \tilde{G}Q^2}\left(-\frac{2\tilde{G}M}{\tilde{r}}+\frac{\tilde{G}Q^2}{\tilde{r}^2}\right),
\end{equation}
which in terms of the original radial coordinate and Newton constant $G$ becomes
\begin{equation}\label{coeffmeric}
    F(r, \epsilon)=1+\frac{r^4}{r^4+2 q Q^2 G g^3(\epsilon)}\left(-\frac{2G M}{r}+\frac{G Q^2 g(\epsilon)}{r^2}\right),
\end{equation}
with $Q$ being the magnetic charge and $q$ the non-minimal coupling constant. We can further redefine those constants and interpret the way the rainbow function appears in the metric as the coupling between the energy of the system probing the space-time and the black hole magnetic charge and nonminimal coupling through the Yang-Mills field. We write then our new magnetic charges and nonminimall couplings as
 \begin{equation}
     \tilde{Q}^2 = Q^2g(\epsilon), \ \tilde{q} = q g(\epsilon)^2. 
 \end{equation}
 This is also consistent with an effectively running gravitational constant and with the general spirit of General Relativity, where space-time tells matter how to move and matter tells space-time how to curve. The rainbow function dependence going from the Newton constant to the field coupling constants just reflects the interaction between matter and gravity and their fundamental connection. From this point on we take $G=1$.

\section{Thermodynamics} \label{thermodynamics-section}
We are now ready to derive the Hawking temperature and other related thermodynamical quantities. It is worth noting that in our case, due to the magnetic charge and nonminimal coupling, even for the case $f = g$ the horizon size and the Hawking temperature will be energy-dependent \cite{Magueijo_2004, Morais2022}. The temperature is defined as
\begin{equation}
    T_H = \frac{1}{2\pi} \lim_{r \rightarrow r_H} \sqrt{\frac{-g^{t t} g^{r r}}{4}\left( \frac{\partial g_{tt}}{\partial r}\right)^2},
\end{equation}
which in terms of our metric function $F(r, \epsilon)$ becomes
\begin{equation}
    T_H = \frac{1}{4\pi}\sqrt{\left(\frac{\partial F(r, \epsilon)}{\partial r}\right)^2}\Bigr|_{\substack{r=r_H}}.
    \label{T_H_diff}
\end{equation}
The horizon equation $g_{t t} = 0$ leads to a mass function given by:
\begin{equation}
   M = \frac{r_H^4+2Q_m^2qg(\epsilon)^3+r_H^2Q_m^2g(\epsilon)}{2r_H^3}.
\end{equation}
For thermodynamic considerations, we make $\epsilon \sim 1/r_H$, where $r_H$ is the radius of the outer horizon. This can be viewed as a limit on the energy scale of the radiated photons given by the uncertainty principle. However, it can also be viewed as a limit on the minimal size of the BH. 

\subsection{Temperature and remnants}

Equation \ref{T_H_diff}, with the substitution of the mass in terms of the horizon radius and the chosen rainbow function, will give us the Hawking temperature as:
\begin{equation}
    T_H=\frac{-6 q Q^2 + (r_H + \lambda)^2 \left[ -Q^2 + r_H \left( r_H + \lambda \right) \right]}{4\pi r_H \left[ 2 q Q^2 + r_H \left( r_H + \lambda \right)^3 \right]}.
\end{equation}

\begin{figure}
    \centering
    \includegraphics[width=0.9\linewidth]{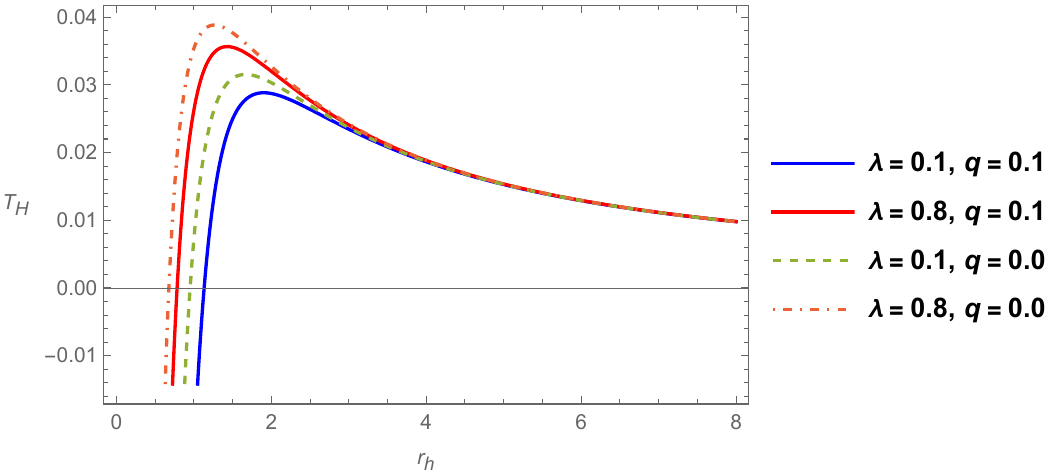}    
    \caption{Hawking temperature as a function of the horizon radius, $r_H$, for some values of the rainbow parameter and the coupling constant, and $Q=1.0$.}
    \label{fig.1}
    \end{figure}
In the Fig \ref{fig.1} we plot the Hawking temperature as a function of $r_H$. The plot illustrates the Hawking temperature \( T_H \) as a function of the event horizon radius \( r_H \). Solid lines represent the regular black hole case (\( q \neq 0 \)), while dashed lines correspond to the singular black hole (\( q = 0 \)). The parameter \( \lambda \) accounts for the effects of Rainbow Gravity.  From the graph, we observe that the Hawking temperature is generally higher for the singular black hole (\( q = 0 \)) compared to the regular case (\( q \neq 0 \)). This suggests that the non-minimal coupling diminishes the black hole's thermal emission. We can also observe that, in both cases, the temperature increases as \( r_H \) diminishes, reaches a maximum, and then decreases until it reaches zero, signaling the formation of a remnant. Notably, in the singular case, the remnant is formed at a smaller event horizon radius compared to the regular black hole. We also note that, in both cases, the temperature increases with the rainbow parameter $\lambda$. Furthermore, as the Rainbow Gravity parameter \( \lambda \) increases, the temperature curves shift, and the remnant forms at an even smaller \( r_H \). This indicates that quantum gravity effects influence both the evaporation process and the final remnant size, with stronger Rainbow Gravity effects leading to smaller remnants. Moreover, in the absence of magnetic charge, no remnants are formed, as expected, since $T_H=(4\pi r_H)^{-1}$.

It should be noted that as $r_H \rightarrow Q$ the BH solution nears its extreme behavior and one should be careful with the subtleties that arise due to the presence of the magnetic charge and the non-Abelian gauge field \cite{Kim2004, Maldacena2021, Ashtekar_2001}. That being said, the temperature analysis \textit{signals} a qualitative behavior that is consistent with what we expect from a regular BH solution and we can confirm the impact of the rainbow parameter through the calculation of other thermodynamic quantities, which we do in the next subsections.

\vspace{-0.5cm}
\subsection{Entropy}

The entropy is given by $S=\int dM/T_H$, with the ADM mass given by
\begin{equation}
M=\frac{r_H}{2} \left\{ 1 + \frac{Q^2 \left[ 2 q + (\lambda + r_H)^2 \right]}{r_H (\lambda + r_H)^3} \right\},
\end{equation}
which yields
\begin{eqnarray}
S=S_0-\frac{2 \pi q Q^2}{r_H^2} + \lambda \pi Q \log \left( \frac{r_H - Q}{r_H + Q} \right),
\end{eqnarray}
up to first order in $\lambda Q$. In this expression, by taking the limit \(Q\to 0 \), we recover the Bekenstein formula \(S= S_0 = \pi r_H^2 \). 

We can observe that the logarithmic term associated with the entropy, which is characteristic of a quantum gravity theory \cite{Meissner_2004,BarberoG_2009, Calmet2021,Xiao2022}, vanish when $\lambda=0$. We also note that the corrected entropy can only be defined when $r_H > Q$. This limit is in agreement with what has been already said regarding the behavior of the solutions when $r_H \rightarrow Q$. As this limit is reached, quantum effects of the gauge field on the vacuum cannot be fully modeled by the usual semiclassical description. Nevertheless, as is the case for the usual magnetically charged Reissner-Nordström  solution, one can still infer the general qualitative behavior of our solutions by simultaneously analyzing the thermodynamic quantities and the regularity of the solutions \cite{Kim2004}.

\subsection{Specific heat}

We now turn our attention to the study of specific heat, given by
\begin{equation}\label{spec.heat}
    C=\frac{dM}{dT}=\frac{dM/dr_H}{dT_H/dr_H}.
\end{equation}
In the context of General Relativity, the specific heat of a Schwarzschild black hole is negative for all values of the horizon radius. This behavior reflects the fact that the temperature of the black hole decreases as the horizon radius increases.
\begin{figure}
    \centering
    \includegraphics[width=0.9\linewidth]{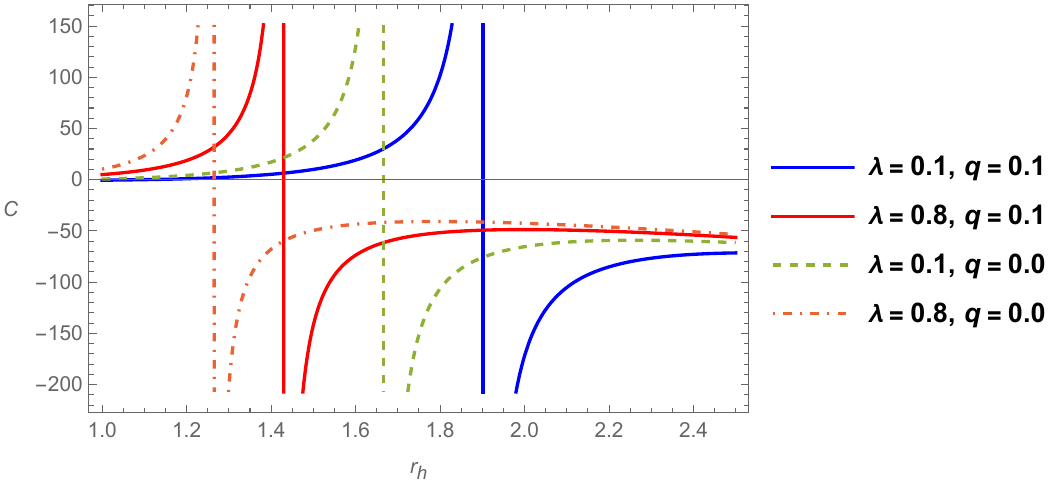}    
    \caption{Specific heat as a function of the horizon radius for selected values of the rainbow parameter and the coupling constant, considering $Q=1.0$.
}
    \label{fig.2}
    \end{figure}
Regarding our black hole solution, the derived expression for the specific heat from Eq. (\ref{spec.heat}) is quite involved. However, in the limit for which $Q=0$, we find the expression for the Schwarzschild black hole, $C=-2\pi r_H^2$. Therefore, we illustrate it in Fig. \ref{fig.2}. The plot reveals, for $q\neq 0$, abrupt phase transitions from negative values of $C$ (indicating local thermodynamic instability) to positive values (signifying local stability). Consequently, a first-order (discontinuous) phase transition occurs. These transitions occur at smaller black hole sizes as the rainbow parameter increases. Moreover, numerical analysis confirms that the remnant state is achieved after the phase transition, which occurs within the stable regime. In other words, as the black hole evaporates, its temperature increases until it reaches a maximum, where the transition takes place, and then it cools down to the remnant state.

We again consider the case with minimal coupling in parallel with the non-minimally coupled one. One should take the positivity of the specific heat as a sign that after a minimum horizon size both cases are potentially stable. However, since this occurs in a region where $r_H$ starts to approach the value of the magnetic charge this should be taken with caution. The fact that with increasing $\lambda$ we get a phase transition for smaller values of the event horizon is consistent with what we would expect if we had ``stronger'' quantum gravity effects. The fact that for $q=0.1$ the phase transition occurs for $1.5 <r_H<2.0$ might indicate that magnetic black holes with nonminimal coupling could reach a stable remnant state before reaching extreme behavior that can induce further instabilities \cite{Kim2004}. We focus our attention now to study the effects of the rainbow function on the emission rate of our system through the calculation of the graybody factor focusing on a nonminmally coupled case with $r_h =  10Q$, thus avoiding the issues that could arise when the BH becomes extreme. Before doing that, a comment on remnants in Rainbow Gravity in general and in our model specifically is required.

\subsection{A comment on remnants in Rainbow Gravity}

The main reason we considered the minimally coupled case in this section was to verify that even for the usual Reissner-Nordström  solution our choice of rainbow functions leads to modified thermodynamics, indicating a path towards the formation of stable remnants. This is in contrast to what has been claimed elsewhere \cite{Morais2022}, where a given choice of rainbow functions was shown to lead to a finite temperature when $r_H = 0$ for some choices of parameters but not to a zero temperature for a finite horizon radius in the case of the modified Schwarszchild metric. We have to point out that for our choice of rainbow functions \ref{g-rainbow} following previous results would lead us to modifications only to the entropy and specific heat functions, but not to the temperature, since in general it is found that $T_{Rainbow} = (f(\epsilon)/g(\epsilon))T_{Schwarszchild}$, in the case of a Schwarzschild BH. In this case, it would be necessary to revaluate the process of BH evaporation taking into consideration not only our different rainbow functions but also the running of the gravitational coupling. It was surprising to find out that this has not been done in previous studies, probably due to the fact that in the Schwarszchild case including an energy dependent $G$ through the transformation $\tilde{G}(\epsilon) = G/g(\epsilon)$, combined with the transformation in the radial coordinate, leads to
\begin{equation}
    \frac{2\tilde{G}(\epsilon)M}{\tilde{r}(\epsilon)} = \frac{2GM}{r},
\end{equation}
which was first shown in \cite{Magueijo_2004}. It seems that in the literature this was taken as a reason to ignore the possible running of the gravitational coupling, but as we shown in our calculations, taking into account an energy dependent $G$ has consequences to other terms of the metric function since they include other types of interaction between the gauge field and gravitation. 

In our case, the running of the gravitational constant is ``transferred'' to the magnetic charge and the nonminimall coupling, but it was only through a consistent solution of the field equations including $\tilde{G}(\epsilon)$ that we were able to find the correct dependence of this terms on the rainbow functions. The specific characteristics of our nonminimally couled Einstein-Yang-Mills BH does not allow us to make general conclusions regarding the fate of remnants in Rainbow Gravity. That being said, our results in this section combined with the graybody calculation and the singularity analysis we will provide in the next sections indicates that the effect of the modified space-time picture we get by the introduction of rainbow functions and an energy dependent gravitational coupling leads to diminishing emission rates and contributes to singularity avoidance. Both effects can be linked to the formation of stable remnants and we take our results to be further indication that there is in fact the possibility of stable remnant formation in Rainbow Gravity. A further study of this considerations will be subject of future work. 


\section{Scalar graybody factor and particle emission rate}\label{gray}

In this section, we calculate the graybody factor associated with the transmission of an outgoing massless scalar field \(\varphi=\varphi(r,t)\) through the effective potential barrier surrounding the black hole. To achieve this, we solve the Klein-Gordon equation in the given spacetime, which is expressed as
\begin{equation}
    \frac{1}{\sqrt{-g}}\partial_{\mu}(\sqrt{-g}g^{\mu\nu}\partial_{\nu}\varphi)=0,
\end{equation}
where $-g$ is the metric determinant. This equation can be reformulated in terms of a Schrödinger-like equation by introducing tortoise coordinates,  leading to the radial equation:
\begin{equation}
    \frac{d^2 \varphi}{dr_{*}^2}+[\omega^2+V_{eff}(r(r_*))]\varphi=0,
\end{equation}
where $\omega$ is the frequency, identified with the particle energy. The tortoise coordinate is defined from $dr_*/dr=F(r)^{-1}$ and the effective potential is written as \cite{Al-Badawi:2023emj}
\begin{equation}
V_{eff}(r)=g(\epsilon)^2\frac{\ell(\ell+1)F(r)}{r^2}+g(\epsilon)^2\frac{F(r)F'(r)}{r}. 
\end{equation}
Here we will consider the mode $\ell=0$.   
\begin{figure}
    \centering
    \includegraphics[width=0.48\linewidth]{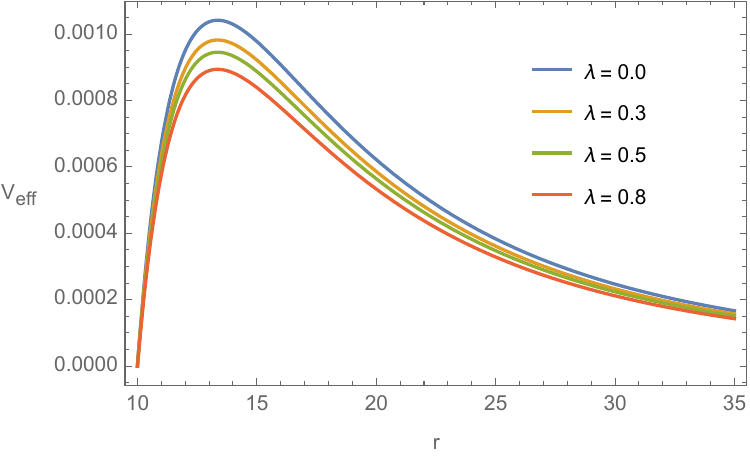}  
    \includegraphics[width=0.48\linewidth]{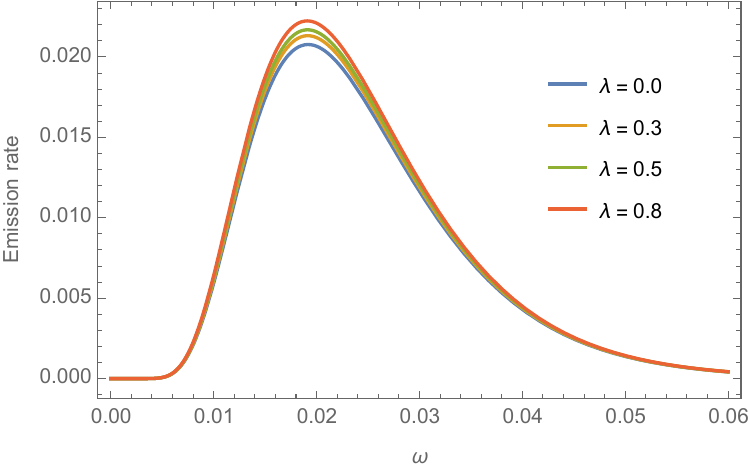}
    \caption{Left panel: Effective potential of a massless scalar field as a function of \(r\) for selected values of the rainbow parameter, with fixed values of \(Q = 1.0\), \(q = 0.1\), \(r_H = 10\), and \(\epsilon = \omega = 0.1\). Right panel: Particle emission rate as a function of the particle energy, using the same set of parameters, except for the energy, which is allowed to vary.}
    \label{fig.3}
    \end{figure}

The graybody factor \(\Gamma_{\ell}^{\omega}\), which quantifies the deviation of a black hole’s radiation from that of an ideal black body, is given by \cite{Al-Badawi:2023emj}
\begin{equation}
    \Gamma_{\ell}^{\omega}=\text{sech}^2\left(\int_{r_H}^{\infty}\frac{V_{eff}}{2\omega}dr_{*}\right),
\end{equation}
and the particle emission rate (number of particles per time unit per frequency unit) is \cite{Calza:2025whq}
\begin{equation}
    \frac{d^2 N}{dtd\omega}=\frac{1}{2\pi}\frac{\Gamma_{0}^{\omega}}{[\exp{\left(\omega/T\right)}-1]},
\end{equation}
for the scalar particles in the zero mode ($\ell=0$), where $T$ is the absolute temperature. 

In the left panel of Figure \ref{fig.3}, we illustrate the effective potential of a massless scalar field under varying values of the rainbow parameter \(\lambda\). As \(\lambda\) increases, the potential barrier decreases, enhancing the particle emission. This aligns with the expectations of rainbow gravity, where energy-dependent modifications to spacetime influence particle dynamics, thereby enhancing quantum effects. Consequently, a reduction in barrier height implies a higher transmission probability for outgoing radiation, leading to an increase in the Hawking temperature. This trend is explicitly confirmed in the right panel of the same figure, reinforcing the impact of rainbow gravity corrections on black hole thermodynamics and the corresponding particle emission.
\vspace{-0.5cm}
\section{Energy conditions and Curvatures} \label{energyandcurvatures}

According to General Relativity, the solutions of Einstein's equations, for example, black holes, compact stars, wormholes, etc. can be solved analytically and/or numerically. However, the theory itself does not determine whether these solutions can be constructed or not since it does not compel the type of matter in these solutions leading to numerous possibilities, including exotic phenomena. One could apply the energy conditions as criteria describing the physical behavior of matter and ruling out the nonphysical ones. On the other hand, one can look for violations of this energy conditions as signs of exotic phenomena or new physics. To generate the energy conditions, the energy-momentum tensor $T_{\mu \nu}$ must contract to the four-vectors. Each type of the four-vectors, for instance, null vector $l^{\mu}$, space-like vector $x^{\mu}$ and time-like vector $t^{\mu}$, provides different energy conditions through different combination of these vectors and energy-momentum tensor. The energy-momentum tensor can be written as 
\begin{eqnarray}
    T_{\mu \nu} = \left( \rho + p_{\perp} \right) t_{\mu} t_{\nu} + p_{\perp} g_{\mu \nu} - (p_{\perp} - p_{r}) x_{\mu} x_{\nu},
\end{eqnarray}
where $t_{\mu} t^{\mu} = 1$, $x_{\mu} x^{\mu} = -1$ and $x^{\mu} t_{\mu} = 0$. $\rho$ is the energy density, $p_r$ is the radial pressure and $p_{\perp}$ is the tangential pressure. In this case, we consider the anisotropic fluid $p_{r} \neq p_{\perp}$. The solutions of the regular black holes are not homogeneous but can still be isotropic. 

Owing to the field equations in Eq.~(\ref{eq:1}) to Eq.~(\ref{eq:3}), the effective density and pressures in both directions are given by
\begin{eqnarray}
    \rho^{eff}&=&-p_r^{eff}=\frac{2 Q^2 g(\epsilon)}{r^4} \left[\frac{1}{2}+\frac{q g^2(\epsilon) F'(r)}{r}-\frac{3 q g^2(\epsilon)F(r)}{r^2}+\frac{3 qg^2(\epsilon)}{r^2}\right]; \label{rho_pr}\\
    p_{\theta}^{eff}&=&p_{\phi}^{eff}= \frac{2 Q^2 g
    (\epsilon)}{r^4} \left[\frac{1}{2}+\frac{1}{2} q g^2(\epsilon) F''(r)-3 q g^2(\epsilon) \left(\frac{F'(r)}{r}-\frac{2 F(r)}{r^2}\right)-\frac{6 q g^2(\epsilon)}{r^2}\right], \label{pt}
\end{eqnarray}
where $F(r)$ and $g(\epsilon)$ come from Eqs. (\ref{F(r)}) and (\ref{g-rainbow}), respectively.
\vspace{-0.5cm}
\subsection{Energy Conditions}
\vspace{-0.5cm}
The energy conditions we will analyze are given as follows.
\begin{itemize}
    \item The null energy condition (NEC) is given by
    \begin{eqnarray}
        T^{eff}_{\mu \nu} l^{\mu} l^{\nu} \geq  0 \rightarrow  \rho^{eff} + p_{i}^{eff} \geq 0,
    \end{eqnarray}
for each $i$ where $i \in \{r, \theta, \phi \}$. The NEC presents the relation of energy density and pressure as measured by a null-like observer, for instance, a photon, must be non-negative. By our derivation of energy density and radial pressure in Eq.~(\ref{rho_pr}), we find that the radial case $\rho^{eff} + p_{r}^{eff}$ is always zero which satisfies the NEC. Since the pressures in $\theta$ and $\phi$ directions are the same as Eq.~(\ref{pt}), we consider the relation $\rho^{eff} + p_{\theta}^{eff}$ as shown in Fig.~\ref{fig_EnPt_r} where we vary the rainbow parameter $\lambda$ from 0.0 (nonminimally copuled GR) to 0.8 and mark the event horizon of each case in the dashed lines. In Figure \ref{fig_EnPt_r} we can see that, at least for the $p_{\theta}^{eff}$ component, the null energy condition is everywhere satisfied, although when $r\rightarrow 0$ the combination of energy density and pressure tends to infinity due to the singularity in the gauge field \cite{BalakinZayats2007}. Because the condition $\rho^{eff} + p_{i}^{eff} \geq 0$ is satisfied in radial and all tangential directions, we conclude that the Einstein-Yang-Mills Regular black holes in rainbow gravity does not violate the null energy condition. 
\begin{figure}[h]
    \centering
    \includegraphics[scale=0.9]{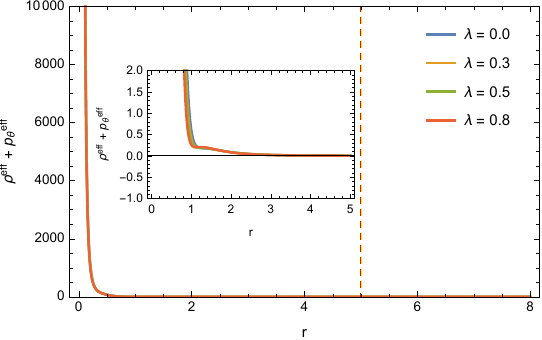}
    \caption{The profile of $\rho^{eff} + p_{\theta}^{eff}$ as distance from the event horizon to $r = 8$ for selected values of the parameter set as $r_H = 5.0, Q = 1.0, q = 0.1$.}
    \label{fig_EnPt_r}
\end{figure}

\item The weak energy condition (WEC) is given by 
\begin{eqnarray}
    T_{\mu \nu}^{eff} t^{\mu} t^{\nu} \geq 0 \rightarrow
    \rho^{eff} + p_{i}^{eff} \geq 0 \text{ and } \rho^{eff} \geq 0,
\end{eqnarray}
for each component $p_i^{eff}$. WEC presents the relations of energy density and pressure as measured by any observer moving through spacetime or time-like path. There are two conditions in WEC; $\rho^{eff} + p_{i}^{eff} \geq 0$ which is NEC and $\rho^{eff} \geq 0$ that the effective energy density must be non-negative. Since NEC for this black hole solutions are satisfied, we investigate the effective energy density in Fig.~\ref{fig_Weak_r}. We can see that when the horizon gets small enough the WEC is not satisfied and we should expect quantum gravity effects to be significant. 
\begin{figure}[h]
    \centering
    \includegraphics[scale=0.9]{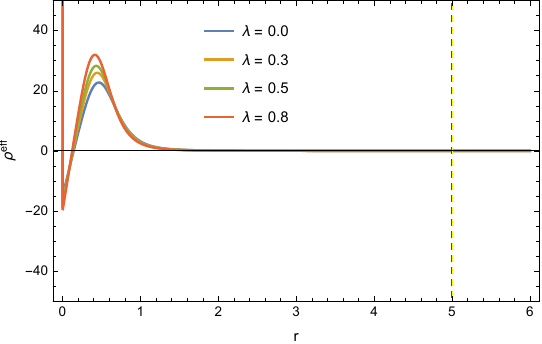}
    \caption{The profile of $\rho^{eff}$ as distance from the event horizon to $r = 10$ for selected values of the parameter set as $r_H = 5.0, Q = 1.0, q = 0.1$.}
    \label{fig_Weak_r}
\end{figure}

\item The dominant energy condition (DEC) will be given by
\begin{eqnarray}
    \rho^{eff} \geq 0 \text{\ and } \rho^{eff} - |p_{i}^{eff}| \geq 0,
\end{eqnarray}
again for each $i$. We observe that the dominant energy condition (DEC), which requires the energy density to be non-negative and the energy flux to remain subluminal, is significantly violated near the origin, with the violation becoming more pronounced for larger values of the rainbow parameter \( \lambda \), as shown in Figure \ref{fig_Dominant}. This suggests that quantum gravity effects enhance the exotic nature of matter in this region. 

\begin{figure}[h]
    \centering
    \includegraphics[scale=0.9]{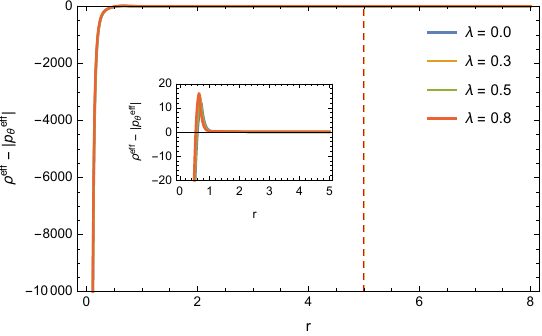}
    \caption{The profile of $\rho^{eff} - |p_{i}^{eff}| \geq 0$ as distance from the event horizon to $r = 10$ for selected values of the parameter set as $r_H = 5.0, Q = 1.0, q = 0.1$.}
    \label{fig_Dominant}
\end{figure}

\item The strong energy condition (SEC)
\begin{equation}
    \rho^{eff} + p_{i}^{eff} \geq 0 \text{ and } \rho^{eff} + p_{r}^{eff} + p_{\theta}^{eff} + p_{\phi}^{eff} \geq 0.
\end{equation}
Likewise, the strong energy condition (SEC), which implies that gravity should remain universally attractive, is also violated within an internal region of the regular black hole, as it must be \cite{Zaslavskii:2010qz}, with the violation intensifying as \( \lambda \) increases, as illustrated in Figure \ref{fig_Strong_r}. This behavior indicates that Rainbow Gravity, by introducing energy-dependent modifications to spacetime close to the Planck scale, amplifies deviations from classical energy conditions, particularly in regions close to the black hole, where quantum effects become most significant.
\end{itemize}

\begin{figure}[h]
    \centering
    \includegraphics[scale=0.9]{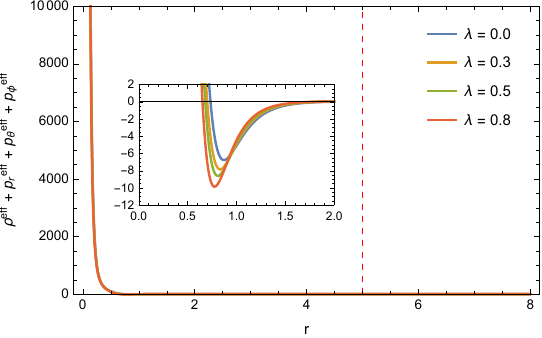}
    \caption{The profile of $\rho^{eff} + p_{r}^{eff} + p_{\theta}^{eff} + p_{\phi}^{eff}$ as distance from the event horizon to $r = 10$ for selected values of the parameter set as $r_H = 5.0, Q = 1.0, q = 0.1$.}
    \label{fig_Strong_r}
\end{figure}

These violations, amplified by the influence of Rainbow Gravity, underscore the intricate interplay between quantum gravity and the internal structure of regular black holes, offering deeper insight into the fundamental modifications induced by high-energy effects.

\subsection{Curvature Scalars}

In this section, we examine the scalar curvatures in order to inspect the regularity of the nonminimally coupled Einstein-Yang-Mills black hole in Rainbow Gravity. At first, the Ricci scalar is given by
\begin{eqnarray}
    R &=& - \frac{g(\epsilon)^2 \left( -2 + 2 F(r) + 4 r F'(r) + r^2 F''(r) \right)}{r^2} \nonumber \\
    &=& -\frac{8 g(\epsilon)^3 q Q^2 \left[6 g(\epsilon)^4 q Q^4 - 20 g(\epsilon)^3 M q Q^2 r - 5 g(\epsilon) Q^2 r^4 + 6 M r^5 \right]}{\left[2 g(\epsilon)^3 q Q^2 + r_H^4\right]^3}.
\end{eqnarray}
At the origin, this curvature is 
\begin{equation}
    R =-\frac{6 (\lambda +r_H)^2}{q r_H^2}.
\end{equation}
The Kretschmann scalar is given by
\begin{eqnarray}\label{kretsch1}
    K = g(\epsilon)^4 \left[ \frac{4(F(r)-1)^2}{r^4} + \frac{4F'(r)^2}{r^2} + F''(r)^2 \right].
\end{eqnarray}
Close to the origin, this quantity can be approximated by
\begin{equation}\label{kretsch2}
    K\approx\frac{20}{9 q^2} + r \left[\frac{1}{9 q^2} - \frac{8}{q^2 r_H} - \frac{8 r_H}{q^2 Q^2} - \frac{8 \lambda}{q^2 Q^2} - \frac{24}{q r_H (r_H + \lambda)^2} \right].
\end{equation}
\begin{figure}[h]
    \centering
    \includegraphics[width= 8.1 cm]{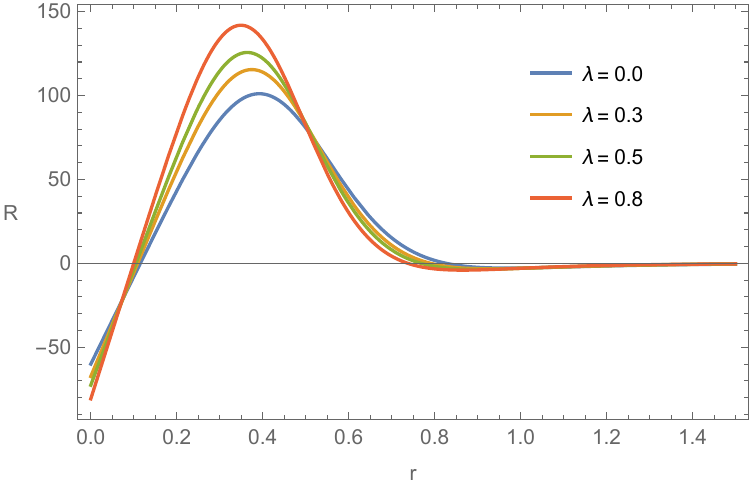}
    \includegraphics[width= 8.1 cm]{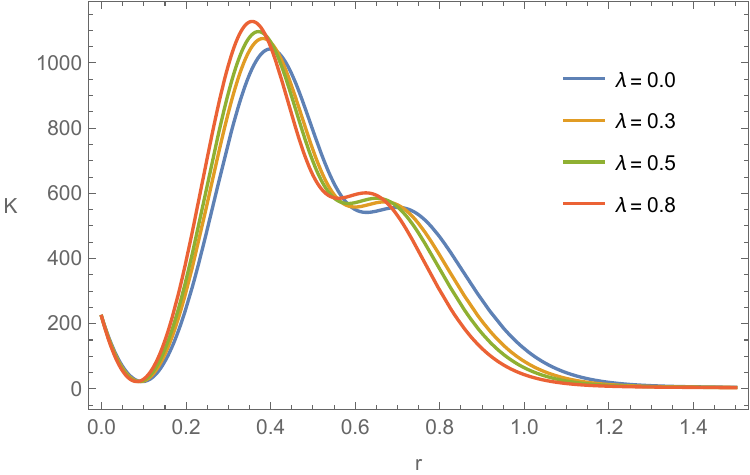}
    \caption{Left panel: Ricci's scalar as a function of $r$, for some values of the rainbow parameter. Right panel: Kretschmann's scalar as a function of $r$ for the same selected values of the rainbow parameter. The remaining parameters are set to $r_H = 5.0$, $Q = 1.0$, and $q = 0.1$.}
    \label{curvatures}
\end{figure}
In the left panel of Figure \ref{curvatures}, we illustrate the Ricci scalar as a function of \( r \) for different values of the rainbow parameter. The absence of divergences confirms the regularity of the solution. Notably, the negative curvature near the origin decreases with increasing \( \lambda \). Additionally, both the local positive maximum and negative minimum of the Ricci scalar increase as this parameter grows. In the right panel of the same figure, we present the Kretschmann scalar, further confirming that the spacetime remains free of singularities. Furthermore, from Eq. (\ref{kretsch2}), we infer that near the origin, as the coupling parameter $q$ tends to zero, the curvature diverges, as expected for a Reissner-N\"{o}rdstron black hole. However, in this region, the positive curvature decreases with increasing \( \lambda \), mitigating the tendency towards singularity formation, provided this parameter is positive definite. Beyond that, the observed local maxima increases with this parameter.

This behavior of spacetime curvatures highlights the influence of quantum gravity corrections within the framework of Rainbow Gravity, where \( \lambda \) modifies the spacetime structure at high energies, significantly altering the curvature profiles.

\section{Shadow of the black holes} \label{shadows}

Although black hole shadow analysis is phenomenologically more relevant for spinning solutions, we use the static one to evaluate the impact of Rainbow Gravity on the shadow size. This should be viewed as an initial step in the direction of comparing the model we analyzed in this work with actual observations \cite{Liu2019pov, Kimet2021,Jiang_2024,LIU2024139052}. We will consider that the system is in a vacuum. Thus, the radius of the shadow is given by  
 \begin{equation}  
    R_{sh} \approx R_o \sin{\alpha_{sh}},  
\end{equation}  
for a distant observer located at \( R_o \) \cite{PERLICK20221}. The angular radius of the shadow \( \alpha_{sh} \) is expressed in the form:  
\begin{equation}  
\sin{\alpha_{sh}} = \frac{\gamma(r_{ph})}{\gamma(R_o)},  
\end{equation}  
where  
\begin{equation}  
\gamma(r) = \sqrt{-\frac{g^{tt}}{g^{\phi\phi}}}. 
\end{equation}  
 The photon sphere radius, \( r_{ph} \), can be determined by solving:  
\begin{equation}  
    \frac{d\gamma^2(r)}{dr} = 0 \quad \text{at} \quad r = r_{ph}.  
\end{equation}  
\begin{figure}
    \centering
    \includegraphics[width=0.55\linewidth]{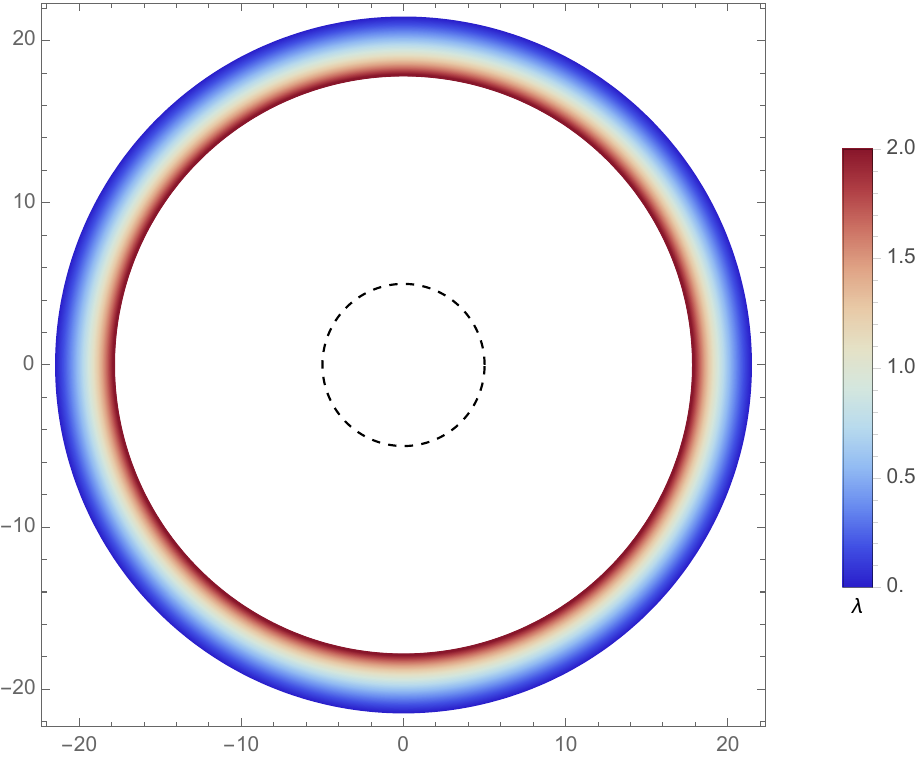}
    \caption{Edges of black hole shadows for a range of rainbow parameter values, spanning from $\lambda=0.0$ (blue) to $\lambda=2.0$ (dark red). The dashed circle represents the event horizon at $r_H=5.0$. Other parameters are fixed at $Q=5.0$ and $q=1.0$.}
    \label{figshadow}
\end{figure}
For our nonminimal Einstein-Yang-Mills black hole solution, where the metric coefficient \( g_{tt} \) is given by Eq. (\ref{coeffmeric}) and \( g_{\phi\phi} = r^2 \) in the equatorial plane. The shadow radius has a complicated expression in terms of the event horizon, the magnetic charge, the nonminimal coupling and the rainbow parameter. Therefore, we illustrate it in Fig. \ref{figshadow}, which depicts the shadow configurations associated with our black hole solutions. Notably, as quantum gravity effects strengthen, the shadow shrinks indicating the possibility of testing our phenomenological rainbow functions and the analyzed black hole solution. 

\section{Conclusions}

In this work, we investigated regular black hole solutions in the framework of Einstein-Yang-Mills theory with non-minimal coupling modified by Rainbow Gravity. By incorporating an energy-dependent modification to the spacetime metric, we analyzed how quantum gravity effects influence the structure, thermodynamics, particle emission, energy conditions, and shadows of these black holes.

From a thermodynamic perspective, we found that Rainbow Gravity. including a consistent running of the gravitational coupling, introduces corrections to the Hawking temperature, entropy, and specific heat. The Hawking temperature increases with the rainbow parameter \( \lambda \), as black hole evaporation is inherently a quantum effect and is consequently amplified by Rainbow Gravity. Additionally, the entropy exhibits a logarithmic correction, a hallmark of quantum gravitational effects \cite{Meissner_2004,BarberoG_2009, Calmet2021,Xiao2022}.  

Through the analysis of specific heat, we identified the occurrence of discontinuous phase transitions when the black hole temperature reaches a maximum, indicating a shift in thermodynamic stability. Notably, the possible existence of black hole remnants is reinforced, with their size decreasing as \( \lambda \) increases. Thus, the stronger the influence of quantum gravity effects, the longer the black hole evaporates before undergoing the phase transition, slowing down this process and ultimately reaching a remnant state with a smaller size, at least from the point of view of the semiclassical analysis. This suggests that the interplay between quantum effects and regularity conditions may prevent the complete evaporation of black holes, potentially offering insights into the resolution of the information loss problem \cite{Chen2015}. As was pointed out in Section \ref{thermodynamics-section}, a complete analysis of the evaporation process for the extreme cases would require considering quantum effects of the gauge fields and a proper discussion on magnetic monopoles which was beyond the scope of our work \cite{Maldacena2021,Ashtekar_2001}. Our focus was to obtain a qualitative analysis of the impact of the rainbow parameter on the thermodynamical behavior of the regular black hole solution and our calculations show that quantum gravity effects do contribute to the stability of remnant states. We briefly discussed the importance of considering the running of the gravitational constant \cite{Magueijo2002, Magueijo2003,Magueijo_2004,Donoghue2012} and the possible implications for the formation of remnants in rainbow gravity, a subject of active debate in the field \cite{Ali2015,JUNIOR2020,Morais2022}. Although our results cannot provide a definite answer on the existence of remnants in Rainbow Gravity, our results show that different rainbow functions and considerations on the running of the Newton constant will generally lead to different physical behaviors. It is important to consider this point in different BH models to better understand the final stages of BH evaporation in the context of Rainbow Gravity. That will be the subject of future works. 

We also analyzed particle emission from the black hole by calculating the graybody factor for a massless scalar field \cite{Al-Badawi:2023emj,Calza:2025whq}. Our findings demonstrate that emission rates increase with the rainbow parameter, highlighting how quantum gravity effects intensify the radiation process, consistent with the observed temperature behavior. 

Our results also shows how that the presence of the rainbow parameter alters the curvature profiles, as observed in the Ricci and Kretschmann scalars. The behavior of this latter near the origin suggests that the rainbow gravity parameter $\lambda$ plays a crucial role in suppressing singularity formation by softening the curvature near the origin, reinforcing the regular nature of the solution. These modifications, therefore, ensure the regularity of the solution while highlighting the role of quantum corrections in high-energy regimes. Furthermore, we analyzed the energy conditions and demonstrated that both the dominant and strong energy conditions are violated in specific regions, with violations intensifying for larger values of \( \lambda \). This behavior indicates that Rainbow Gravity enhances exotic matter effects in the black hole. 

Finally, we explored the impact of Rainbow Gravity on the black hole shadow radius, showing that it decreases as quantum gravity effects strengthen. This result suggests that observational constraints on black hole shadows may serve as a potential probe for testing Rainbow Gravity models in astrophysical scenarios.

Overall, our findings reinforce the idea that Rainbow Gravity provides a compelling avenue for studying quantum gravitational corrections to black hole physics by showing that the rainbow functions can modify thermodynamical quantities, enhance particle emissions, affect the regularity of the solutions and directly impact observations through the shadow size. Future investigations could extend this analysis to rotating black holes, as well as explore potential observational signatures that could distinguish Rainbow Gravity effects from other quantum gravity models. There should be special attention given to the impact of the running of the gravitational constant in different analysis of BH models in the context of Rainbow Gravity.  

This work shows that the study of BH thermodynamics continues to be a rich subject to explore the limits of our theories and investigate the impact of new proposed theories on previously studied models. The current age of precision cosmology and astrophysics motivates us to explore diverse theoretical scenarios that could potentially solve some long standing issues of modern physics, like the singularity problem, the origins of the dark sector and the possible breaking of Lorentz symmetry at high energies. By analyzing this regular BH model in the context of Rainbow Gravity we provided a discussion that connects this issues and contributes with the ongoing debates on the formation of remnants and their possible role in explaining Dark Matter. 

\section*{Acknowledgments}
FBL is funded by Fundação Cearense de Apoio ao Desenvolvimento Científico e Tecnológico (FUNCAP) and by  Conselho Nacional de Desenvolvimento Científico e Tecnológico (CNPq), grant number 305947/2024-9. CRM is funded by Conselho Nacional de Desenvolvimento Científico e Tecnológico (CNPq), under the grant 308268/2021-6.

\bibliography{references.bib} 
\end{document}